\documentclass[10pt,final,journal,twocolumn]{IEEEtran}
\ifCLASSINFOpdf
\else
\fi
\usepackage{color}
\usepackage{cite}
\usepackage[cmex10]{amsmath}
\usepackage{amsthm}
\usepackage{algorithm}
\usepackage{array}
\usepackage{algorithmic}
\usepackage{stfloats}
\usepackage{mathrsfs}
\usepackage{subfigure}
\usepackage{bm}
\usepackage{multicol,multienum}
\usepackage[pdftex]{graphicx}
\usepackage{longtable}
\usepackage{rotating}
\usepackage{multirow}
\begin{document}
%

\title{D2D Communications Underlaying Wireless Powered Communication Networks}

\author{Haichao Wang, Jinlong Wang,~\IEEEmembership{Senior Member,~IEEE,} \\
Guoru Ding,~\IEEEmembership{Senior Member,~IEEE,} and Zhu Han,~\IEEEmembership{Fellow,~IEEE}

\thanks{Copyright (c) 2015 IEEE. Personal use of this material is permitted. However, permission to use this material for any other purposes must be obtained from the IEEE by sending a request to pubs-permissions@ieee.org.

This work is supported by the National Natural Science Foundation of China (Grant No. 61501510), Natural Science Foundation of Jiangsu Province (Grant No. BK20150717), China Postdoctoral Science Foundation Funded Project (Grant No. 2016M590398), and Jiangsu Planned Projects for Postdoctoral Research Funds (Grant No. 1501009A). The research is partially supported by US MURI, NSF CNS-1717454, CNS- 1731424, CNS-1702850, CNS-1646607.}

\thanks{H. Wang, J. Wang, and G. Ding are with College of Communications Engineering, Army Engineering University of PLA, Nanjing 210007, China (email: whcwl0919@sina.com, wjl543@sina.com, dr.guoru.ding@ieee.org). G. Ding is also with National Mobile Communications Research Laboratory, Southeast University, Nanjing 210096, China.}

\thanks{Z. Han is with the University of Houston, Houston, TX 77004 USA (e-mail:zhan2@uh.edu), and also with the Department of Computer Science and Engineering, Kyung Hee University, Seoul, South Korea.}

}

%
%

\IEEEpeerreviewmaketitle
\maketitle
\begin{abstract}
In this paper, we investigate the resource allocation problem for D2D communications underlaying wireless powered communication networks, where multiple D2D pairs harvest energy from a power station equipped with multiple antennas and then transmit information signals simultaneously over the same spectrum resource. The aim is to maximize the sum throughput via joint time scheduling and power control, while satisfying the energy causality constraints. The formulated non-convex problem is first transformed into a nonlinear fractional programming problem with a tactful reformulation. Then, by leveraging D.C. (difference of two convex functions) programming, a suboptimal solution to the non-convex problem is obtained by iteratively solving a sequence of convex problems. Simulation results demonstrate that the proposed scheme works well in different scenarios and can significantly improve the system throughput compared with the-state-of-the-art schemes.
\end{abstract}


%
\IEEEpeerreviewmaketitle

\section{Introduction}
\IEEEPARstart{A}{dvocated} by the dual use of radio frequency signals, wireless energy transfer (WET) has attracted much attention for improving the system energy efficiency\cite{CST}. In this context, simultaneous wireless information and power transfer (SWIPT)\cite{SWIPT} and wireless powered communication networks (WPCNs)\cite{TDMA1,TDMA2,TDMA4} have been extensively studied in the literature. Moreover, since electromagnetic waves decay quickly over distance, energy beamforming is generally designed to achieve efficient WET\cite{EB}. In WPCNs, a power station (PS) transfers wireless energy to some low-power users with a single antenna due to the hardware constraint. Afterwards, the users transmit information signals with the harvested energy. For the multiple users scenario, the signals are transmitted typically based on time division multiple access (TDMA) as in\cite{TDMA1,TDMA2,TDMA4}. However, the spectrum efficiency can be greatly improved with appropriate interference management methods by allowing multiple users to transmit signals simultaneously.

In this paper, we consider the D2D communications underlaying WPCNs, where the low-power D2D transmitters (D2D-Txs) with one antenna must harvest energy from the PS equipped with multiple antennas before transmitting signals. In the considered scenario, all D2D-Txs transmit information signals simultaneously over the same spectrum resource. Our aim is to maximize the sum throughput via joint time scheduling and power control, while satisfying the energy causality constraints. For solving the formulated non-convex nonlinear problem, we develop a throughput maximization algorithm, where the problem is first transformed into a nonlinear fractional programming problem with a tactful reformulation. Then, an iterative algorithm is designed to address the equivalent problem by leveraging D.C. (difference of two convex functions) programming. In-depth simulations are conducted to evaluate the throughput performance under various system parameter configurations.

The rest of this paper is organized as follows. In Section II, we illustrate the system model and formulate the optimization problem. Then, we develop a throughput maximization algorithm in Section III. In Section IV, we present simulation results to verify the effectiveness of the proposed algorithm. Finally, we conclude the paper in Section V.
\section{System Model and Problem Formulation}
Consider a WPCN with a PS equipped with $M$ antennas and $N$ low-power D2D pairs denoted by $\mathcal{N}= \{ 1, 2, ..., n, ..., N\}$. The D2D pair carries a single antenna due to the size and cost constraints, such as the sensor node\cite{Sensor}. With no embedded energy supply, each D2D-Tx first harvests energy from wireless signal transmitted by the PS (i.e., WET phase). Then, they utilize the harvested energy to transmit information signals to their intended receivers in the wireless information transmission (WIT) phase. According to the harvest-then-transmit protocol, in each block denoted by $T$, the first $\tau_0T$ amount of time, $0 \leq \tau_0 \leq 1$, is assigned to harvest energy for all D2D pairs, while the followed $\tau_1T$ amount of time in the same block is assigned to transmit information signals. Followed by\cite{TDMA4}, we consider a normalized unit block time $T = 1$ in the sequel without loss of generality. Then, there is $\tau_0 + \tau_1 \leq 1$. All the users considered in this paper operate on a single spectrum band\cite{WZ}.

In the WET phase, the $M \times 1$ transmitted signal is given by $\sqrt {{p_{PS}}} {\bf{w}}$, where $p_{PS}$ is the transmit power of the PS, and the beamformer ${\bf{w}}$ is designed to improve the energy transfer efficiency and subject to ${\left\| {\bf{w}} \right\|_2} = 1$. Let $\bf{h}_n$ represent the $M$ dimensional energy transfer channel vector between the PS and $n$-th D2D-Tx. The energy harvested from the noise can be ignored since the noise power is usually much smaller than that of the PS. Therefore, the energy harvested at the $n$-th D2D-Tx is given by
${E_n} = \eta {\tau _0}{p_{PS}}{\left| {{\bf{h}}_n^H{\bf{w}}} \right|^2}$ with energy conversion efficiency $0 < \eta  < 1$. We apply the asymptotically optimal energy beamforming proposed in\cite{beamform} as
\begin{align}
\label{Beamform}
{\bf{w}} = \sum\limits_{n = 1}^N {\sqrt {{\varsigma _n}} \frac{{{{\bf{h}}_n}}}{{{{\left\| {{{\bf{h}}_n}} \right\|}_2}}}},
\end{align}
where $\{ \varsigma_n \}$ controls the energy allocation among multiple D2D pairs and $\sum\nolimits_{n = 1}^N {{\varsigma _n}}  = 1$. In this paper, equal weight is designed for all D2D pairs, i.e., ${\varsigma _n} = {1 \mathord{\left/
 {\vphantom {1 N}} \right.
 \kern-\nulldelimiterspace} N}$. Notice that other energy beamforming schemes as invetigated in\cite{beamform} can be employed and the proposed algorithm still works.

Denote $g_{n,n}$ as the channel power gain from the $n$-th D2D-Tx to its receiver. The channel power gain of the interference link from the $n$-th D2D-Tx to the $k$-th D2D receiver (D2D-Rx) is denoted by ${\tilde g_{n,k}}$. Since all D2D-Txs transmit information signals simultaneously over the same spectrum resource, the signal to interference plus noise ratio at the $n$-th D2D-Rx is as follows:
\begin{align}
\label{SINR}
{\gamma _n} = \frac{{{p_n}{g_{n,n}}}}{{\sum\nolimits_{m \ne n}^N {{p_m}{{\tilde g}_{m,n}}}  + {\sigma ^2}}},
\end{align}
where $p_n$ is the transmit power of $n$-th D2D-Tx and $\sigma ^2$ is the noise power. The achievable throughput at the $n$-th receiver in bits/second/Hz is thus given by
\begin{align}
{r_n} = \tau_1 {\log _2}\left( {1 + {\gamma _n}} \right).
\end{align}
Intuitively, a D2D pair closer to the PS can harvest more energy in short time and vice versa, which potentially results in various energy constraints for different D2D pairs. To character this difference, the transmit power and time are jointly optimized here. The aim is to maximize the sum throughput of all D2D pairs via time scheduling and power control, while satisfying the energy causality constraints. Thus, the optimization problem can be formulated as the following:
\begin{align}
\label{eq:pr1}
P1: &\mathop {\max }\limits_{{\tau _0},{\tau _1},\left\{ {{p_n}} \right\}} {\tau _1}\sum\limits_{n = 1}^N {{{\log }_2}\left( {1 + {\gamma _n}} \right)} \nonumber\\
&s.t. ~C1: \tau_1 (p_n +p_c) \le  \eta {\tau _0}{p_{PS}}{\left| {{\bf{h}}_n^H{\bf{w}}} \right|^2},  ~~~\forall n,\nonumber\\
&~~~~~C2: \tau_0 + \tau_1 \leq 1, \nonumber\\
&~~~~~C3: 0 \le \tau_0, \tau_1 \le 1,\nonumber\\
&~~~~~C4: {p_n} \geq 0,~~~\forall n,
\end{align}
where $p_c$ represents the non-ideal circuit power consumption (e.g., AC/DC converter, analog amplifier, and processor)\cite{TDMA1,TDMA2}. $C1$ guarantees that the consuming energy by any D2D-Tx cannot exceed its harvested energy. $C2, C3$ and $C4$ are the time and power control constraints. However, the feasible region of $C1$ is non-convex, which means that the standard convex optimization methods cannot be used to efficiently solve this problem\cite{Convex}. Even if the time lengths of WET and WIT have been fixed, the investigated problem is still non-convex and hard to be addressed. In the next section, we propose an efficient throughput maximization algorithm by exploiting the problem structure.
\section{Throughput Maximization Algorithm for D2D Communications}
The optimal solution to the problem (\ref{eq:pr1}) is generally difficult to be obtained since there are multiple local optima due to the non-convex nonlinear property. To this end, the formulated non-convex problem is first transformed into a nonlinear fractional programming problem with a tactful reformulation. Then, an iterative algorithm is designed to solve the equivalent problem by leveraging D.C. programming.

The time utilization is illustrated by the following Lemma.

\emph{Lemma 1:} The optimal solution to the problem (\ref{eq:pr1}) is achieved if and only if all the time is used, i.e., ${\tau_0} + {\tau_1} = 1$.

\begin{IEEEproof}
To prove Lemma 1, we assume that $\left\{ {{\tau_0}^\prime ,{\tau_1}^\prime, \left\{ {{p_n}^\prime } \right\}} \right\}$ is the optimal solution satisfying ${\tau_0}^\prime + {\tau_1}^\prime = \xi_1 < 1$ and ${\tau_1} ^\prime= \delta {\tau_0} ^\prime = {\delta  \mathord{\left/
 {\vphantom {\delta  {\left( {1 + \delta } \right)}}} \right.
 \kern-\nulldelimiterspace} {\left( {1 + \delta } \right)}} \xi_1$, which means that there is remaining time available denoted by $\xi_2 = 1 - \xi_1$. If we can find a feasible solution to the optimization problem (\ref{eq:pr1}) in the remaining time, it demonstrates that the system throughput can also be improved. In other words, the solution $\left\{ {{\tau_0}^\prime ,{\tau_1}^\prime, \left\{ {{p_n}^\prime } \right\}} \right\}$ is not the optimal solution.

The remaining time $\xi_2$ can also be divided into two parts according to the ratio $\delta$. The first ${1 \mathord{\left/
 {\vphantom {1 {\left( {1 + \delta } \right)}}} \right.
 \kern-\nulldelimiterspace} {\left( {1 + \delta } \right)}} \xi_2$ amount of time of the remaining time $\xi_2$ is used to harvest energy, while the remaining is to transmit information signals. Since $\left\{ {{\tau_0}^\prime ,{\tau_1}^\prime, \left\{ {{p_n}^\prime } \right\}} \right\}$ is a feasible solution, there must be
 \begin{align}
\frac{\delta }{{1 + \delta }}{\xi _1}\left( {{p_n} + {p_c}} \right) \le \frac{1}{{1 + \delta }}{\xi _1}\eta {p_{PS}}{\left| {{\bf{h}}_n^H{\bf{w}}} \right|^2}.
 \end{align}
 Then, the following constraint also holds true:
 \begin{align}
\frac{\delta }{{1 + \delta }}{\xi _2}\left( {{p_n} + {p_c}} \right) \le \frac{1}{{1 + \delta }}{\xi _2}\eta {p_{PS}}{\left| {{\bf{h}}_n^H{\bf{w}}} \right|^2}.
 \end{align}
This means that the solution $\left\{ {{\tau_0}^{\prime \prime },{\tau_1}^{\prime \prime }, \left\{ {{p_n}^{\prime \prime }} \right\}} \right\}$ is a feasible solution in the remaining time with ${\tau _0}^{\prime \prime } = {1 \mathord{\left/
 {\vphantom {1 {\left( {1 + \delta } \right){\xi _2}}}} \right.
 \kern-\nulldelimiterspace} {\left( {1 + \delta } \right){\xi _2}}}$, ${\tau _1} ^{\prime \prime }= {\delta  \mathord{\left/
 {\vphantom {\delta  {\left( {1 + \delta } \right){\xi _2}}}} \right.
 \kern-\nulldelimiterspace} {\left( {1 + \delta } \right){\xi _2}}}$, and ${p_n}^{\prime \prime }={p_n}^\prime$. So the remaining time can be used to improve the system throughput, which contradicts the assumption. The Lemma 1 has been proved.
\end{IEEEproof}

Based on Lemma 1, the constraint $C1$ in problem (\ref{eq:pr1}) can be transformed as follows:
\begin{align}
{\tau_1} \left( {{p_n} + {p_c}} \right) - \left( {1 - {\tau_1}} \right) \eta {p_{PS}}{\left| {{\bf{h}}_n^H{\bf{w}}} \right|^2} \le 0, ~~~\forall n.
\end{align}
It is non-convex with respect to ${\tau_1}$ and $p_n$, which hinders the application of standard convex optimization techniques. Although it can be transformed into a convex function by geometric programming\cite{Convex}, the resulting objective function will be much more complicated. To this end, replace ${\tau_1} = {1 \mathord{\left/
 {\vphantom {1 t}} \right.
 \kern-\nulldelimiterspace} t}$ and $t \geq 1$. The convenience of this replacement will be shown later. Thus, the constraint $C1$ in problem (\ref{eq:pr1}) can be rewritten as
\begin{align}
\left( {{p_n} + {p_c}} \right) - t\eta {p_{PS}}{\left| {{\bf{h}}_n^H{\bf{w}}} \right|^2} + \eta {p_{PS}}{\left| {{\bf{h}}_n^H{\bf{w}}} \right|^2} \le 0,~~~\forall n.
\end{align}
This alternation makes the constraint $C1$ become linear. With this reformulation, the optimization problem (\ref{eq:pr1}) is equivalent to
\begin{align}
\label{P1v}
&\mathop {\max }\limits_{{t},\left\{ {{p_n}} \right\}} ~ \frac{1}{t}\sum\limits_{n = 1}^N {\log_2 \left( {1 + \frac{{{p_n}{g_{n,n}}}}{{\sum\nolimits_{m \ne n}^N {{p_m}{{\tilde g}_{m,n}}} + {\sigma ^2}}}} \right)} \nonumber\\
&s.t.~ C1: \left( {{p_n} + {p_c}} \right) - t\eta {p_{PS}}{\left| {{\bf{h}}_n^H{\bf{w}}} \right|^2} + \eta {p_{PS}}{\left| {{\bf{h}}_n^H{\bf{w}}} \right|^2} \le 0,\forall n, \nonumber\\
&~~~~~C2:t \ge 1,\nonumber\\
  &~~~~~C3:{p_n} \ge 0{\kern 1pt} ,{\kern 1pt} {\kern 1pt} {\kern 1pt} {\kern 1pt} {\kern 1pt} {\kern 1pt} \forall n.
\end{align}

It can be observed that the non-convex constraints are transformed into convex functions with a tactful reformulation. Moreover, the optimization problem (\ref{P1v}) can be seen as a nonlinear fractional programming\cite{Nfp}. Therefore, we try to search an optimal solution to the problem (\ref{eq:pr1}) by solving the equivalent problem (\ref{P1v}).

Denote ${q^ * }$ as the optimal solution of the considered problem (\ref{P1v}), which is given by
\begin{align}
\label{Pt}
{q^ * } = \frac{{\sum\nolimits_{n = 1}^N {{R_n}\left( {{\left\{ {{p_n}} \right\}^ * }} \right)} }}{{{t^ * }}} = \mathop {\max }\limits_{t,\left\{ {{p_n}} \right\}} \frac{{\sum\nolimits_{n = 1}^N {{R_n}\left( \left\{ {{p_n}} \right\} \right)} }}{t},
\end{align}
where
\begin{align}
\label{Pt}
{R_n} \left( \left\{ {{p_n}} \right\} \right)\; = {\log _2}\left( {1 + \frac{{{p_n}{g_{n,n}}}}{{\sum\nolimits_{m \ne n}^N {{p_m}{{\tilde g}_{m,n}}} + {\sigma ^2}}}} \right).
\end{align}

The following Lemma 2 provides a guidance on designing an iterative approach to solve the problem (\ref{P1v}).

\emph{Lemma 2}: The optimal solution is achieved if and only if
\begin{align}
\label{Pt11}
\mathop {\max }\limits_{t, {\left\{ {{p_n}} \right\} } } ~\sum\limits_{n = 1}^N {{R_n}\left( \left\{ {{p_n}} \right\} \right)}  - {q^ * }t = \sum\limits_{n = 1}^N {{R_n}\left( {{\left\{ {{p_n}} \right\}^ * }} \right)}  - {q^ * }{t^ * } = 0.
\end{align}

Lemma 2 can be proven by following a similar approach as in\cite{Nfp}. It is shown that the original fractional form in problem (\ref{P1v}) can be transformed into a subtractive form with an equivalent solution, which indicates that an iterative algorithm can be designed to solve this problem. Specifically, initializing from a given $q$, we should solve a sequence of following problems:
\begin{align}
\label{Pt14}
&\mathop {\max }\limits_{t,\left\{ {{p_n}} \right\}} ~f\left( {t,q,\left\{ {{p_n}} \right\}} \right) = \sum\limits_{n = 1}^N {{R_n}\left( \left\{ {{p_n}} \right\} \right)}  - qt\nonumber\\
&s.t.~C1,C2,C3~\text{in}~ \left( 9 \right).
\end{align}

The overall procedure for solving the optimization problem (\ref{P1v}) is presented in \textbf{Algorithm 1}. The convergence to the optimal solution is guaranteed.

\textbf{\emph{Theorem 1}}: As long as the number of iterations is sufficiently large, the proposed algorithm will eventually approach the optimal solution.

\begin{IEEEproof}
See Appendix A for details.
\end{IEEEproof}

\begin{algorithm}[t]
\caption{Proposed throughput maximization algorithm}
\begin{algorithmic}[1]
\STATE Initialize the parameter $q$
\STATE \textbf{Repeat}
     \STATE ~~~~Solve the problem (\ref{Pt14}) for a given $q$ to obtain $\left( {t',{\left\{ {{p_n}} \right\}^\prime } } \right)$
     \STATE ~~~~~~Set $q = {{\sum\nolimits_{n = 1}^N {{R_n} \left( {\left\{ {{p_n}} \right\}^\prime } \right)} } \mathord{\left/
 {\vphantom {{\sum\limits_{n = 1}^N {{R_n}^\prime\left( {\left\{ {{p_n}} \right\}^\prime } \right)} } {t'}}} \right.
 \kern-\nulldelimiterspace} {t'}}$
\STATE \textbf{Until} some termination conditions are met
\STATE Return ${t^ * } = t'$, $p_n^ *  = {p_n}^\prime $, ${q^ * } = {{\sum\nolimits_{n = 1}^N {{R_n}\left( {\left\{ {{p_n}} \right\}^ * } \right)} } \mathord{\left/
 {\vphantom {{\sum\limits_{n = 1}^N {{R_n}\left( {{\bf{p}}^ * } \right)} } {{t^ * }}}} \right.
 \kern-\nulldelimiterspace} {{t^ * }}}$
\end{algorithmic}
\end{algorithm}

Although we have designed a framework to efficiently solve the problem (\ref{P1v}), the optimization problem (\ref{Pt14}) is hard to be solved due to the non-convexity of the objective function. The optimal solution to this problem cannot be obtained so far. In the sequel, we develop an iterative algorithm to get a suboptimal solution to the problem (\ref{Pt14}).

Denote the following concave functions
\begin{align}
\label{first-order}
&w_n\left( \left\{ {{p_n}} \right\} \right)={\log _2}\left( {\sum\nolimits_{m = 1}^N {{p_m}{{\tilde g}_{m,n}}}  + {\sigma ^2}} \right),\nonumber\\
&v_n\left( \left\{ {{p_n}} \right\}\right) = {\log _2}\left( {\sum\nolimits_{m \ne n}^N {{p_m}{{\tilde g}_{m,n}}} + {\sigma ^2}} \right).
\end{align}
The objective function in problem (\ref{Pt14}) can be expressed as
\begin{align}
f\left( {t,q,\left\{ {{p_n}} \right\}} \right) = \sum\limits_{n=1}^N {{w_n}\left( \left\{ {{p_n}} \right\} \right)}  - \sum\limits_{n=1}^N {{v_n}\left( \left\{ {{p_n}} \right\} \right)}  - qt.
\end{align}
It can be observed that the objective function is the difference of two concave functions. A series of non-decreasing solutions can be obtained by iteratively optimizing the lower bound of objective function, which is given by the following Lemma 3.

\emph{Lemma 3}: Given ${{\left\{ {{p_n}} \right\}^\prime }}$, the following function is a tight lower bound of the objective function in problem (\ref{Pt14}):
\begin{align}
\label{lowbound}
&f\left( {t,q,\left\{ {{p_n}} \right\},{\left\{ {{p_n}} \right\}^\prime }} \right) = \sum\limits_{n=1}^N {{w_n}\left( \left\{ {{p_n}} \right\} \right)}  - \sum\limits_{n=1}^N {{v_n}\left( {{\left\{ {{p_n}} \right\}^\prime }} \right)}\nonumber\\
 &~~~~~~~~~- \sum\limits_{n=1}^N {\left\langle {\nabla {v_n}\left( {{\left\{ {{p_n}} \right\}^\prime }} \right),\left( {\left\{ {{p_n}} \right\} - {\left\{ {{p_n}} \right\}^\prime }} \right)} \right\rangle }  - qt,
\end{align}
where the $l$-th component of the $\nabla v_n\left( {{\left\{ {{p_n}} \right\}^\prime }} \right)$ is given by
\begin{align}
\nabla {v_n}\left( {{\left\{ {{p_n}} \right\}^\prime }} \right) = \frac{1}{{\ln 2}}\frac{{{{\tilde g}_{l,n}}}}{{\sum\nolimits_{m \ne n}^N {{p_m}^\prime {{\tilde g}_{m,n}}} + {\sigma ^2}}}.
\end{align}

\begin{IEEEproof}
Since $v_n\left( \left\{ {{p_n}} \right\} \right)$ is concave, based on the first-order condition of a concave function, we have $v_n\left( \left\{ {{p_n}} \right\} \right) \leq v_n\left( {{\left\{ {{p_n}} \right\}^\prime }} \right) + \left\langle {\nabla v_n\left( {{\left\{ {{p_n}} \right\}^\prime }} \right),\left( {\left\{ {{p_n}} \right\} - {\left\{ {{p_n}} \right\}^\prime }} \right)} \right\rangle$\cite{Convex}. Thus, $f\left( {t,q,\left\{ {{p_n}} \right\}} \right) \ge f\left( {t,q,\left\{ {{p_n}} \right\}, \left\{ {{p_n}} \right\}^\prime } \right)$. Moreover, if $\left\{ {{p_n}} \right\} = {\left\{ {{p_n}} \right\}^\prime }$, there is $f\left( {t,q,\left\{ {{p_n}} \right\}} \right) = f\left( {t,q,\left\{ {{p_n}} \right\},{\left\{ {{p_n}} \right\}^\prime }} \right)$. So $f\left( {t,q,\left\{ {{p_n}} \right\},{\left\{ {{p_n}} \right\}^\prime }} \right)$ provides a tight lower bound of function $f\left( {t,q,\left\{ {{p_n}} \right\}} \right)$.
\end{IEEEproof}

According to Lemma 3, an iterative algorithm can be developed to solve the optimization problem (\ref{Pt14}). In particular, initializing from a given ${{\left\{ {{p_n}} \right\}^\prime }}$, we can iteratively solve the following convex problem using standard convex optimization techniques:
\begin{align}
\label{eq:innerP}
&\mathop {\max }\limits_{t,\left\{ {{p_n}} \right\}} f\left( {t,q,\left\{ {{p_n}} \right\},{\left\{ {{p_n}} \right\}^\prime }} \right)\nonumber\\
&s.t.~C1,C2,C3 ~\text{in}~ \left( 9 \right).
\end{align}
There are many convex optimization techniques and they have been widely investigated\cite{Convex}. Due to the page limit, we omit the procedure of solving problem (\ref{eq:innerP}). Finally, the overall procedure for solving the optimization problem (\ref{Pt14}) is presented in \textbf{Algorithm 2}.

\emph{Lemma 4}: The resulting values of Algorithm 2 at each iteration are non-decreasing, and the convergence is guaranteed.

\begin{IEEEproof}
Let ${\left\{ {{p_n}} \right\}^k}$ be the solution at $k$-th iteration. Since $f\left( {t,q,{{\left\{ {{p_n}} \right\}}^{k + 1}},{{\left\{ {{p_n}} \right\}}^k}} \right)$ provides a lower bound for $f\left( {t,q,{{\left\{ {{p_n}} \right\}}^{k + 1}}} \right)$, there is
\begin{align}
f\left( {t,q,{{\left\{ {{p_n}} \right\}}^{k + 1}}} \right) \ge f\left( {t,q,{{\left\{ {{p_n}} \right\}}^{k + 1}},{{\left\{ {{p_n}} \right\}}^k}} \right).
\end{align}
Moreover, because ${{{\left\{ {{p_n}} \right\}}^{k + 1}}}$ is the optimal solution at the $(k+1)$th iteration, we have
\begin{align}
&f\left( {t,q,{{\left\{ {{p_n}} \right\}}^{k + 1}},{{\left\{ {{p_n}} \right\}}^k}} \right) = \mathop {\max }\limits_{t,\left\{ {{p_n}} \right\}} f\left( {t,q,\left\{ {{p_n}} \right\},{{\left\{ {{p_n}} \right\}}^k}} \right)\nonumber\\
& \ge f\left( {t,q,{{\left\{ {{p_n}} \right\}}^k},{{\left\{ {{p_n}} \right\}}^k}} \right).
\end{align}
Therefore, there is
\begin{align}
&f\left( {t,q,{{\left\{ {{p_n}} \right\}}^{k + 1}}} \right) \ge f\left( {t,q,{{\left\{ {{p_n}} \right\}}^k},{{\left\{ {{p_n}} \right\}}^k}} \right) = f\left( {t,q,{{\left\{ {{p_n}} \right\}}^k}} \right).
\end{align}
It can be observed that the resulting values are non-decreasing at each iteration. Further, it must be upper bounded by the optimal value of (\ref{Pt14}). Thus, the convergence is guaranteed.
\end{IEEEproof}
\begin{algorithm}[t]
\caption{SCA for solving problem in (\ref{P1})}
\begin{algorithmic}[1]
\STATE Input the value $q$ and initialize $(t, \left\{ {{p_n}} \right\})$
\STATE \textbf{Repeat}
\STATE ~~~~~Solve the problem (\ref{eq:innerP}) via standard convex optimization techniques and obtain the optimal solution $(t^ \circ, \left\{ {{p_n}} \right\}^ \circ)$
\STATE ~~~~Update ${\left\{ {{p_n}} \right\}^{k+1}} = {\left\{ {{p_n}} \right\}^ \circ }$ and ${t^{k + 1}} = {t^ \circ}$
\STATE \textbf{Until} some termination conditions are met
\end{algorithmic}
\end{algorithm}

\emph{Complexity Analysis:} The proposed throughput maximization algorithm contains two-layers iterations. The outer-layer iteration is the update procedure defined by (\ref{Pt14}). The inner-layer iteration is to acquire a lower bound by solving the D.C. programming problem at given $q$. The convex problem (\ref{eq:innerP}) can be solved with the complexity of $O((N+1)^3)$ by standard convex optimization techniques, such as interior point method, where $N$ is the number of D2D pairs. Assume that $K, L$ are the computations in the two-layer iterations. The total complexity can be roughly estimated as $O(KL(N+1)^3)$.

Notably, we consider a centralized network in this paper, where the PS performs the proposed algorithm. The channel state information can be estimated by the PS according to the channel reciprocity\cite{beamform}.
\section{Simulations and Discussions}
In this section, we perform in-depth simulations to evaluate the performance of the proposed algorithm in a $50 \times 50$ m area, where multiple D2D pairs are randomly located and the maximum distance between D2D-Tx and D2D-Rx is $D=10$ m. The channel power gain is modeled as ${g} =10^{-3} \rho^2{d^{ - \alpha }}$\cite{TDMA2,beamform}, where $\rho^2$ is an exponentially distributed random variable with unit mean, $d$ is the distance between the transmitter and receiver, and ${\alpha }=3$ represents the path-loss exponent. Unless specified otherwise, the bandwidth is 1 MHz and noise power spectral density is $-170$ dBm/Hz. The transmit power and number of antennas are 1 W and 10 for the PS, respectively. The energy conversion efficiency and circuit power consumption are 0.5 and 0.1 $\mu$W. In all simulations, $q=1$ is set to start the algorithm and all results are averaged over 100 realizations.

\begin{figure}[!t]
\centering {\includegraphics[width=80mm]{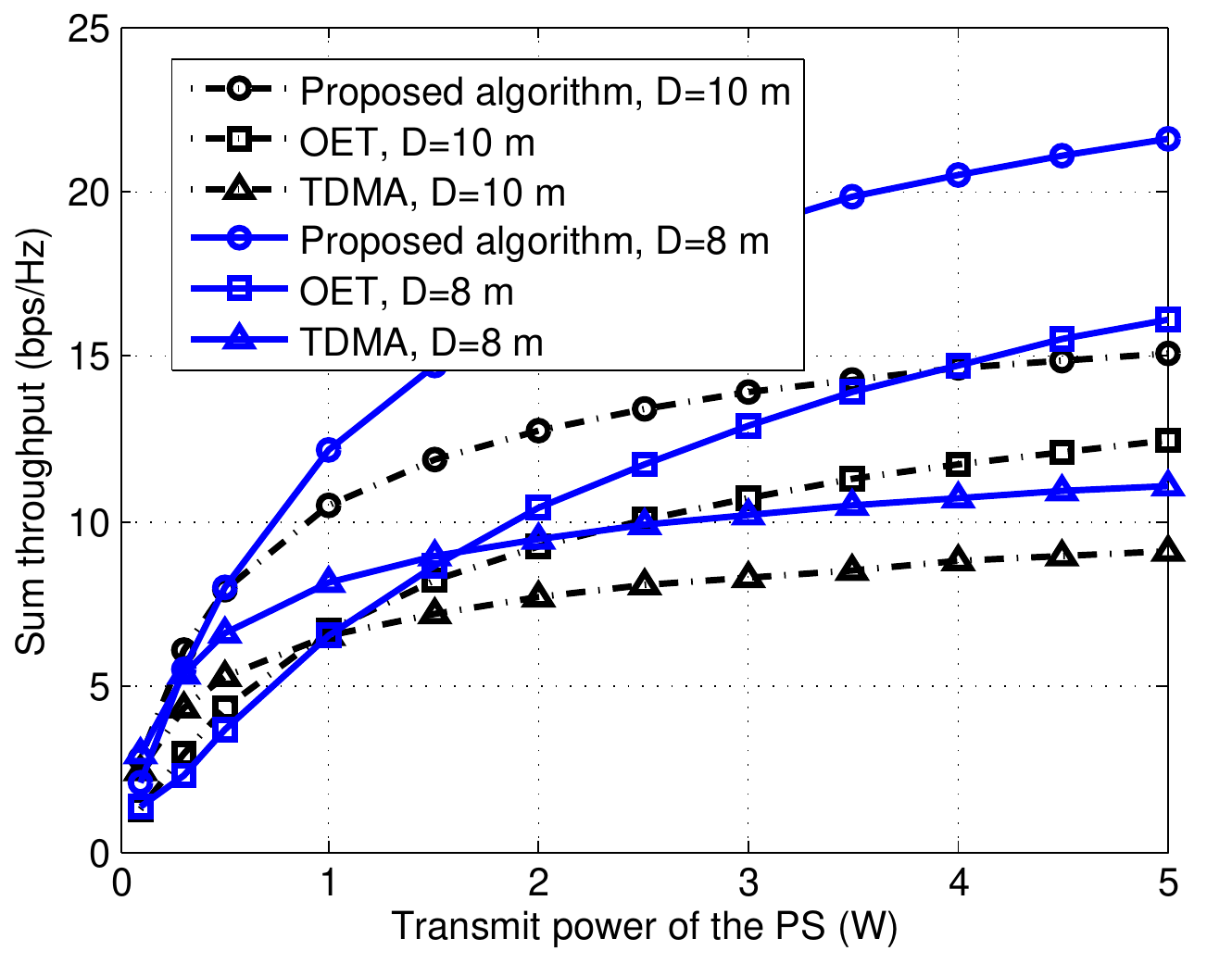}}
\caption{The throughput performance comparison of different schemes versus the transmit power of the PS.}
\label{fig:performance}
\end{figure}
The sum throughput versus the transmit power of the PS is shown in Fig. \ref{fig:performance}. For comparison, we also provide an energy transfer scheme without beamforming, namely the omnidirectional energy transfer (OET)\cite{ICC}, and a TDMA-based algorithm where multiple users harvest energy and then transmit information signals based on TDMA. It can be observed that the proposed algorithm outperforms the OET and TDMA-based algorithm in all cases. Furthermore, we can observe that the growth rate gradually becomes slower as the transmit power increases. This is due to the fact that the mutual interference among D2D pairs dominates the system with sufficiently large transmit power. In addition, The final sum throughput would be better if the maximum distance $D$ between D2D-Tx and D2D-Rx is reduced. The reason is that smaller maximum distance results in better channel state.

\begin{figure}[!t]
\centering {\includegraphics[width=80mm]{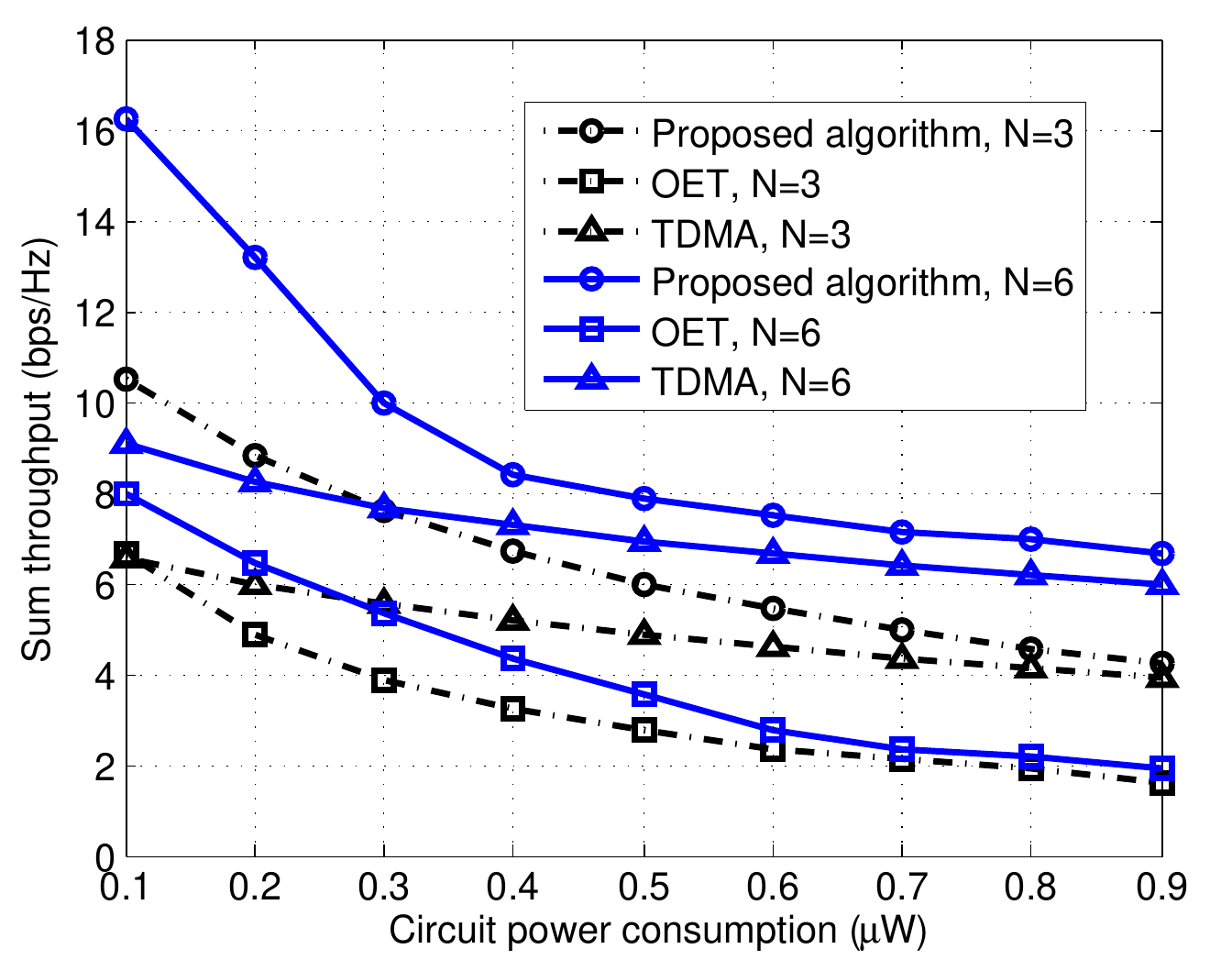}}
\caption{The throughput performance comparison versus circuit power consumption.}
\label{fig:Power}
\end{figure}
In Fig. \ref{fig:Power}, the sum throughput is plotted against the circuit power consumption $p_c$. It can be observed that the sum throughput decreases with an increasing circuit power consumption. Meanwhile, the throughput gain between the proposed algorithm and the TDMA-based algorithm is smaller. The reason is that D2D pairs have little energy for information transmission and some D2D pairs may even stop working since they have not enough energy. The impact of number of antennas is further investigated and simulation results are shown in Table I. The plot confirms the intuition that the sum throughput grows as more antennas are added at the PS since more antennas can make use of the spatial resource to improve diversity gain.
\renewcommand\arraystretch{1}
\begin{table}[!t]
\centering
\caption{The sum throughput versus the number of antennas}
\begin{tabular}{|c|c|p{0.7cm}<{\centering}|p{0.8cm}<{\centering}|p{0.8cm}<{\centering}|p{0.8cm}<{\centering}|p{0.8cm}<{\centering}|p{0.8cm}<{\centering}|}
\hline
\multicolumn{2}{|c|}{\multirow{2}*{Settings} }&\multicolumn{6}{c|}{Number of antennas ($M$)}\\
\cline{3-8}
\multicolumn{2}{|c|}{}&1&2&3&5&10&15\\
\hline
\multirow{3}*{\rotatebox{90}{Number}}&3&3.2021&4.5291&5.5410&7.1474&10.0983&12.3943\\
\cline{2-8}
&6&4.9674&7.2365&8.5764&11.1300&16.0135&19.2880\\
\cline{2-8}
&9&6.9566&9.8967&11.5237&15.0312&19.5413&23.0866\\
\hline
\end{tabular}
\end{table}

Table II shows the WIT time for different settings, which characterizes the time split between energy harvesting and data transmission. The data transmission time grows with an increasing transmit power of the PS. This is due to the fact that the D2D-Txs can harvest more energy with high transmit power, which reduces the energy harvesting time. On the other hand, more D2D pairs not always results in an increasing WIT time since the mutual interference becomes serious, where some D2D pairs experienced serious interference do not work. It can be also observed that the value ($N=6, p_{PS}=2$ W) is smaller, which results from locally optimal points of non-convex optimization.
\begin{table}[!t]
\centering
\caption{The WIT time for different settings}
\begin{tabular}{|c|c|c|c|c|c|c|c|}
\hline
\multicolumn{2}{|c|}{\multirow{2}*{Settings} }&\multicolumn{6}{c|}{Transmit power of the PS $p_{PS}$ (W)}\\
\cline{3-8}
\multicolumn{2}{|c|}{}&1&2&3&4&5&6\\
\hline
\multirow{3}*{\rotatebox{90}{Number}}&3&0.4971&0.6641&0.7356&0.7764&0.8033&0.8223\\
\cline{2-8}
&6&0.6594&0.4349&0.7358&0.8527&0.8584&0.8808\\
\cline{2-8}
&9&0.6602&0.7758&0.8217&0.8464&0.8525&0.8733\\
\hline
\end{tabular}
\end{table}

\section{Conclusion}
In this paper, we investigated the resource allocation scheme for D2D communications underlaying WPCNs, where the harvest-then-transmit protocol was employed. We tried to maximize the sum throughput of all D2D pairs while satisfying the energy causality constraints. The considered joint time scheduling and power control problem was formulated as a non-convex optimization problem and then it was transformed into a nonlinear fractional programming problem with a tactful reformulation. By leveraging D.C. programming, a suboptimal solution of the non-convex problem can be obtained by iteratively solving a sequence of convex problems. In-depth simulations were conducted to validate the effectiveness of the proposed algorithm.

\section*{Appendix A}
\section*{The proof of Theorem 1}

The following proof follows the proof of algorithm convergence in paper\cite{Nfp}. For notational convenience, denote ${\bf{p}} = \left\{ {{p_n}} \right\}$ and $F\left( {q'} \right) = \mathop {\max }\limits_{t,{\bf{p}}} \sum\nolimits_{n = 1}^N {{R_n}\left( {\bf{p}} \right)}  - q't$. For any feasible solution $\left( {t',{\bf{p'}}} \right)$ and $q' = {{\sum\nolimits_{n = 1}^N {{R_n}\left( {{\bf{p'}}} \right)} } \mathord{\left/
 {\vphantom {{\sum\nolimits_{n = 1}^N {{R_n}\left( {{\bf{p'}}} \right)} } t}} \right.
 \kern-\nulldelimiterspace} t^\prime }$, there is
$
F\left( {q'} \right) = \mathop {\max }\limits_{t,{\bf{p}}} \sum\limits_{n = 1}^N {{R_n}\left( {\bf{p}} \right)}  - q't \ge \sum\limits_{n = 1}^N {{R_n}\left( {{\bf{p'}}} \right)}  - q't' = 0,
$
which means that $F\left( {q'} \right) \ge 0$ always holds true for any feasible solution $\left( {t',{\bf{p'}}} \right)$. Furthermore, let $\left( {t',{\bf{p'}}} \right)$ and $\left( {t'',{\bf{p''}}} \right)$ as the optimal solutions for $F\left( {q'} \right)$ and $F\left( {q''} \right)$, respectively. Then, there is
$
\label{ZM3}
F\left( {q'} \right) = \sum\limits_{n = 1}^N {{R_n}\left( {{\bf{p'}}} \right)}  - q't' > \sum\limits_{n = 1}^N {{R_n}\left( {{\bf{p''}}} \right)}  - q't''.
$
If $q'' > q'$,
$
F\left( {q'} \right) > \sum\limits_{n = 1}^N {{R_n}\left( {{\bf{p''}}} \right)}  - q't''
> \sum\limits_{n = 1}^N {{R_n}\left( {{\bf{p''}}} \right)}  - q''t''= \mathop {\max }\limits_{t,{\bf{p}}} \sum\limits_{n = 1}^N {{R_n}\left( {\bf{p}} \right)}  - q''t = F\left( {q''} \right).
$
So $F\left( {q'} \right)$ is a strictly monotonic decreasing function.

Denote $\left( {{t^k},{{\bf{p}}^k}} \right)$  as the optimal solution at the $k$-th iteration and the according value is ${q_k} \ne {q^ * }$. From the iterative algorithm, we know ${q^{k + 1}} = {{\sum\nolimits_{n = 1}^N {{R_n}\left( {{{\bf{p}}^k}} \right)} } \mathord{\left/
 {\vphantom {{\sum\nolimits_{n = 1}^N {{R_n}\left( {{{\bf{p}}^k}} \right)} } {{t^k}}}} \right.
 \kern-\nulldelimiterspace} {{t^k}}}$. Then, we have
$ F\left( {{q^k}} \right) = \sum\limits_{n = 1}^N {{R_n}\left( {{{\bf{p}}^k}} \right)}  - {q^k}{t^k} = {t^k}\left( {{q^{k + 1}} - {q^k}} \right) > 0.$
Because ${t^k } > 0 $, so there is ${q^{k + 1}} > {q^k}$. we can show that as long as the number of iterations is large
enough,  $F\left( {{q^k}} \right)$  will eventually approach zero with the increasing $q^k$ since $F\left( {q'} \right)$ is a strictly decreasing function.
 \hfill \IEEEQED

\ifCLASSOPTIONcaptionsoff
  \newpage
\fi


\begin{thebibliography}{3}
\bibitem{CST}
X. Lu, P. Wang, D. Niyato, D. I. Kim, and Z. Han, ``Wireless networks with RF energy harvesting: A contemporary survey,'' \emph{IEEE Commun. Surv. \& Tutor.}, vol. 17, no. 2, pp. 757-789, Nov. 2015.
\bibitem{SWIPT}
K. Huang and E. Larsson, ``Simultaneous information and power transfer for broadband wireless systems,'' \emph{IEEE Trans. Signal Process.}, vol. 61, no. 23, pp. 5972-5986, Dec. 2013.
\bibitem{TDMA1}
Z. Hadzi-Velkov, I. Nikoloska, G. K. Karagiannidis, and T. Q. Duong, ``Wireless networks with energy harvesting and power transfer: Joint power and time allocation,'' \emph{IEEE Signal Process. Lett.}, vol. 23, no. 1, pp. 50-54, Jan. 2016.
\bibitem{TDMA2}
S. Pejoski, Z. Hadzi-Velkov, T. Q. Duong, and C. Zhong, ``Wireless powered communication networks with non-ideal circuit power consumption,'' \emph{IEEE Commun. Lett.}, vol. 21, no. 6, pp. 1429-1432, June 2017.
\bibitem{TDMA4}
H. Ju and R. Zhang, ``Throughput maximization in wireless powered communication networks,'' \emph{IEEE Trans. Wireless Commun.}, vol. 13, no. 1, pp. 418-428, Jan. 2014.
\bibitem{EB}
S. Bi, C. K. Ho, and R. Zhang, ``Wireless powered communication: Opportunities and challenges,'' \emph{IEEE Commun. Mag.}, vol. 53, no. 4, pp. 117-125, Apr. 2015.
\bibitem{Sensor}
W. Huang, H. Chen, Y. Li, and B. Vucetic, ``On the performance of multi-antenna wireless-powered communications with energy beamforming,'' \emph{IEEE Trans. Veh. Technol.}, vol. 65, no. 3, pp. 1801-1808, Mar. 2016.
\bibitem{WZ}
W. Zhao and S. Wang, ``Resource sharing scheme for device-to-device communication underlaying cellular networks,'' \emph{IEEE Trans. Commun.}, vol. 63, no. 12, pp. 4838-4848, Dec. 2015.
\bibitem{beamform}
G. Yang, C. K. Ho, R. Zhang, and Y. L. Guan, ``Throughput optimization for massive MIMO systems powered by wireless energy transfer,'' \emph{IEEE J. Sel. Areas Commun.}, vol. 33, no. 8, pp. 1640-1650, Aug. 2015.
\bibitem{Convex}
S. Boyd and L. Vandenberghe, \emph{Convex Optimization}. Cambridge University Press, 2004.
\bibitem{Nfp}
W. Dinkelbach, ``On Nonlinear Fractional Programming,'' \emph{Manage. Sci.}, vol. 13, no. 7,  pp. 492-498, Mar. 1967.
\bibitem{ICC}
H. Wang, G. Ding, J. Wang, L. Wang, T. A. Tsiftsis, and P. K. Sharma, ``Resource allocation for energy harvesting-powered D2D communications underlaying cellular networks,'' in \emph{Proc. IEEE International Conference on Communications (ICC)}, May 2017, pp. 1-6.
\end{thebibliography}
\end{document}